\documentclass{edp-conf}
\usepackage{graphicx}
\begin{document}
\def\la{\buildrel<\over\sim}
\def\ga{\buildrel>\over\sim}

\TitreGlobal{SF2A 2005}

\title{EVOLUTION OF GALACTIC FIELD Be STARS}

\author{Zorec, J.}\address{Institut d'Astrophysique de Paris, UMR7095 CNRS, 
Universit\'e Pierre \& Marie Curie}
\author{Fr\'emat, Y.}\address{Royal Observatory of Belgium}
\author{Cidale, L.$^3$}\address{Facultad de Ciencias Astron\'omicas y 
Geof\'\i sicas, Universidad de La Plata, Argentina}

\runningtitle{Evolution of Be Stars}

\index{Zorec, J.}
\index{Fr\'emat, Y.}
\index{Cidale, L.}

\maketitle

\begin{abstract}Galactic field Be stars were studied by taking into account 
the effects induced by the fast rotation on their fundamental parameters. 
Fractional ages $\tau/\tau_{\rm MS}$ ($\tau_{\rm MS}$ = time spent in the MS)
against stellar mass reveal that: a) Be stars spread over the whole interval 
$0<\tau/\tau_{\rm MS}<1$; b) the Be phenomenon in massive stars 
($M>12M_{\odot}$) is present at smaller age ratios than for less massive stars 
($M<12M_{\odot}$); c) there is a lack of Be stars with $M<7M_{\odot}$ in the 
first half of the MS. Low mass fast rotators ($M<7M_{\odot}$), called Bn stars, 
could be ``becoming" Be stars.  
\end{abstract}

\section{Sample and Method}

 We studied 97 field galactic bright Be stars using the BCD system (Chalonge 
\& Divan 1956, Zorec \& Briot 1991) to avoid perturbations on the fundamental
parameters due to the circumstellar disc. These parameters were treated for 
rotational effects (Fr\'emat et al. 2005). From the observed parameters, 
called {\it apparent}, we obtain first the {\it parent non-rotating 
counterparts} ({\it pnrc}) which represent ho\-mo\-lo\-gous stars without 
rotation. To enter the evolutionary tracks, we transform then the {\it pnrc} 
into {\it averaged} parameters over the whole stellar surface.
   
\section{Results and Conclusions}

 Figure 1a) shows the distribution of points ($\tau/\tau_{\rm MS},M/M_{\odot}$)
obtained using the original or $apparent$ fundamental parameters and the 
evolutionary tracks without rotation (Schaller et al. 1992). The plotted error
bars correspond to measurement uncertainties. Figure 1b) shows the same type 
of distribution, but where parameters were corrected for rotational effects 
assuming rigid rotation with $\Omega/\Omega_{\rm crit} =$ 0.9 (Fr\'emat et al.
2005) and models of stellar evolution with rotation calculated by Meynet \& 
Maeder (2000) with ZAMS equatorial velocity $V_{\rm ZAMS} =$ 300 km~s$^{-1}$.
In both diagrams of Fig. 1, points spread over the whole interval of age 
fractions $0\la$ $\tau/\tau_{\rm MS}\la1$, which suggest that the Be 
phenomenon may appear at whatever stage of the stellar evolution on the MS 
evolution phase. There is, however, a difference between the diagrams where 
rotation is taken into account and where it is not. If we were not aware 
of rotational effects, Fig. 1a would suggest that 86\% of stars are above the 
$\tau/\tau_{\rm MS} =$ 0.5 limit. Figure 1b shows, however, when fast rotation 
of Be stars is taken into account, the fraction of stars in our sample above 
$\tau \simeq$ $0.5\tau_{\rm MS}$ slumps to 62\%. When we separate the stars 
into $massive$ ($M\ga12M_{\odot}$) and $less\ massive$ ones ($M\la12M_{\odot}$)
another important result appears: we see in Fig. 1b that the Be phenomenon in 
massive stars tends to appear on average at smaller $\tau/\tau_{\rm MS}$ age 
fractions than in the less massive stars. There is also a striking lack of 
low-mass Be stars ($M\la7M_{\odot}$) in $\tau/\tau_{\rm MS}<1/2$. These 
distributions can be due to: $i$) higher mass-loss rates in massive objects, 
which reduce the surface fast rotation and inhibit the Be phenomenon rapidly; 
$ii$) circulation time scales to transport angular momentum from the core to 
the surface, which are longer the lower the stellar mass. Some of low-mass 
fast rotators, classified as Bn stars, could then be {\bf ``becoming"} Be 
stars.

\begin{figure}[h]
\centering
\includegraphics[height=5.9cm,width=9.2cm]{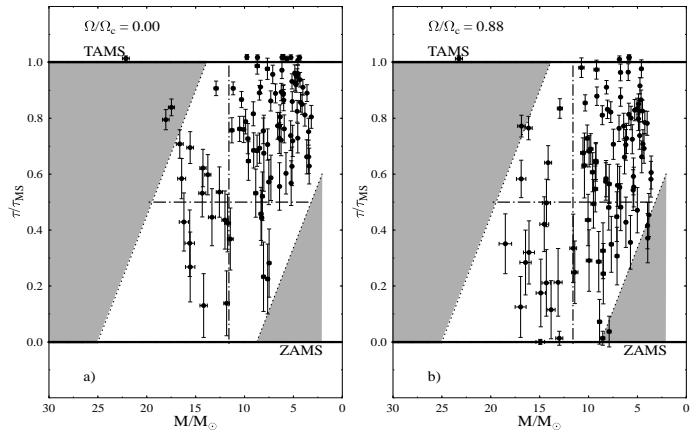}
\caption{Age ratios $\tau/\tau_{\rm MS}$ of the studied Be stars against 
the mass. a) Parameters derived using evolutionary tracks without rotation;
b) parameters derived using evolutionary tracks with rotation}
\end{figure}

\end{document}